\documentclass[twocolumn,preprintnumbers,amsmath,amssymb,showpacs]{revtex4}
\usepackage{epsfig}
\begin{document}
\setlength{\voffset}{1.0cm}
\title{Integrating out the Dirac sea: Effective field theory approach \\to exactly solvable four-fermion models}
\preprint{FAU-TP3-07/08}
\author{Felix Karbstein\footnote{Electronic address:  felix@theorie3.physik.uni-erlangen.de}}
\author{Michael Thies\footnote{Electronic address: thies@theorie3.physik.uni-erlangen.de}}
\affiliation{Institut f\"ur Theoretische Physik III,
Universit\"at Erlangen-N\"urnberg, D-91058 Erlangen, Germany}
\date{\today}
\begin{abstract}
We use 1+1 dimensional large $N$ Gross-Neveu models as a laboratory to derive microscopically effective
Lagrangians for positive energy fermions only. When applied to baryons, the Euler-Lagrange equation for
these effective theories assumes the form of a non-linear Dirac equation. Its solution reproduces the full
semi-classical results including the Dirac sea to any desired accuracy. Dynamical effects from the Dirac
sea are encoded in higher order derivative terms and multi-fermion interactions with perturbatively calculable,
finite coefficients. Characteristic differences between models with discrete and continuous chiral symmetry 
are observed and clarified. 
\end{abstract}
\maketitle
\section{Introduction}
Exactly solvable models in quantum field theory which bear any resemblance to the real world are extremely
rare. One example is provided by the Gross-Neveu model family \cite{C1}, 1+1 dimensional four-fermion interaction
models with $N$ flavors and Lagrangian 
\begin{equation}
{\cal L}=\bar{\psi}\left({\rm i}\partial \!\!\!/-m_0\right)\psi + \frac{g^2}{2} \left[(\bar{\psi}\psi)^2+\lambda (\bar{\psi}
{\rm i}\gamma_5\psi)^2\right],
\label{A1}
\end{equation}
where $\lambda=0$ or 1, and flavor indices are suppressed ($\bar{\psi} \psi=\sum_{k=1}^N\bar{\psi}_k\psi_k$ etc.).
Depending on the value of $\lambda$, these models feature discrete ($\psi \to \gamma_5 \psi$, $\lambda=0$) or
continuous ($\psi \to \exp({\rm i}\alpha \gamma_5) \psi$, $\lambda=1$) chiral symmetry, possibly broken by the bare 
mass term $\sim m_0$. To avoid confusion we shall refer to the first model as Gross-Neveu (GN$_2$), the 2nd one as
Nambu--Jona-Lasinio (NJL$_2$) model. From the point of view of strong interaction
physics, they are most useful in the 't~Hooft limit ($N\to \infty, Ng^2={\rm const.}$) to which we 
stick in the following.
Rather than repeating all the well-known attractive features of these models, we refer the reader to some
pertinent review articles \cite{C2,C3,C4} and the references therein.  

The Lagrangian (\ref{A1}) is known to possess multi-fermion bound states analogous to baryons or
baryonium states in hadron physics. They were first constructed by Dashen, Hasslacher and Neveu (DHN)
in the massless GN$_2$ model \cite{C5} and by Shei in the massless NJL$_2$ model \cite{C6}. The semi-classical
method developed by these authors may be rephrased equivalently as a Dirac-Hartree-Fock calculation \cite{C7,C8}. Here,
one solves the first-quantized, time-independent Dirac equation
\begin{equation}
\left( - \gamma_5 {\rm i}\partial_x + \gamma^0 S +{\rm i}\gamma^1 P \right) \psi_{\alpha}=E_{\alpha}\psi_{\alpha}
\label{A2}
\end{equation}
for single particle orbits with label $\alpha$ subject to self-consistency conditions for scalar and
pseudoscalar mean fields,
\begin{eqnarray}
S & = & m_0 - Ng^2 \sum_{\alpha}^{\rm occ}\bar{\psi}_{\alpha}\psi_{\alpha},
\nonumber \\
P & = &  - \lambda Ng^2 \sum_{\alpha}^{\rm occ} \bar{\psi}_{\alpha}{\rm i} \gamma_5 \psi_{\alpha}.
\label{A3}
\end{eqnarray}
A major challenge arises from the fact that the Dirac sea must be included in the sum over occupied states in (\ref{A3}).
This problem was solved in Refs.~\cite{C5,C6} by inverse scattering methods together with a careful subtraction
of bound state and vacuum energies. At about the same time the proposal was made to solve these models
``classically", neglecting the Dirac sea and including only the discrete valence level \cite{C9,C10}. This reduces the full 
problem to a non-linear 
Dirac equation, which has been solved in closed analytical form for both the GN$_2$ and NJL$_2$ models.
It appeared that the classical fermionic solution was useful for the GN$_2$ model but completely
failed in the NJL$_2$ case \cite{C6} for poorly understood reasons.
The Dirac sea also caused considerable difficulties in attempts to base nuclear physics on field
theoretic models like the ($\sigma,\omega$) model \cite{C11}, so that mean field calculations were typically
done without the sea.

In the present work we reconsider the role of the Dirac sea in multi-fermion bound states of
Gross-Neveu models. From a general field theoretic point of view, it should be possible to ``integrate out"
the Dirac sea and derive an effective Lagrangian for (positive energy) valence fermions only. In the large $N$ limit,
it would then indeed be sufficient to solve the classical Euler-Lagrange equation 
of this effective theory for fermions. From such a point of
view the above mentioned ``no-sea" calculations may be re-interpreted as follows: The authors
assumed that the effective Lagrangian is the same as the original Lagrangian (\ref{A1}),
except that the bare
parameters $m_0$ and $g^2$ are replaced by an effective mass and coupling constant. In the course
of this paper we will confront this implicit assumption with our results for an effective action
derived from the underlying field theory.

Let us mention that although the solvable models used here are restricted to 1+1 dimensions, the
technique applied in deriving the effective Lagrangian is not. It might be useful also in higher dimensional
(mean field type) theories. The advantage of developing the methods in the context
of Gross-Neveu models is the 
fact that the exact bound states are known analytically. This enables
us to test our effective action quantitatively and make sure that the expansion is consistent to a given
order in some small parameter. Since the derivation of ${\cal L}_{\rm eff}$ is not completely straightforward due
to the necessity of resummations, this has turned out to be quite helpful indeed.

This paper is organized as follows. In Sect.~II we derive the effective Lagrangian for the GN$_2$ model
in the three lowest orders of a systematic expansion valid for small filling fraction and/or large bare fermion mass
and check it against the DHN baryon.
In Sect.~III we take up the NJL$_2$ model. Due to additional complications, we will content ourselves with the leading
order effective action here and test it against the Shei bound state. We end with a short summary and  
conclusions in Sect.~IV.

\section{Gross-Neveu model (GN$_2$)}
Our aim is to account for the effects of the Dirac sea by means of an effective theory of positive energy fermions only.
Since we do not use the path integral but work canonically, we first
have to explain what we mean by ``integrating out the Dirac sea". We start with the massless GN$_2$ model
described by the bare Lagrangian 
\begin{equation}
{\cal L}=  \bar{\psi} {\rm i} \partial \!\!\!/ \psi + \frac{g^2}{2} (\bar{\psi}\psi)^2
\label{p1}
\end{equation}
and focus on the DHN baryons. They are 
characterized by a discrete positive energy level occupied with $n\le N$ fermions, all negative energy
levels being completely filled.  
The presence of extra fermions polarizes the Dirac sea, 
distorting the single particle wave functions.
Ideally, one would like to fully integrate out the Dirac sea. 
We have not been able to do this. Anyway, the full effective theory will invariably be non-local and therefore of less practical use.
We therefore look for an expansion parameter which would enable us to derive an ``almost local" effective action, consisting
of a polynomial in the chiral condensate $\bar{\psi}\psi$ and its first few derivatives.
From the DHN baryon we know that for small filling fraction $\nu=n/N$ of the valence level, the self-consistent scalar potential $S(x)$ 
differs from the physical (vacuum) fermion mass $m$ by a weak and slowly varying potential only. This suggests to use
$\nu$ as expansion parameter and to restrict oneself to small filling fraction where the HF potential becomes soft.  

The following derivation of the effective action is tailored to the HF approach to which we now turn.
In the vacuum, a fermion mass is generated dynamically as shown graphically in Fig.~1. 
\begin{figure}
\begin{center}
\epsfig{file=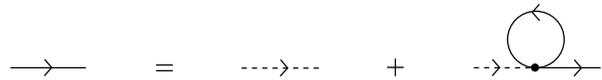,width=8.0cm,angle=0}
\end{center}
\caption{Dyson equation for fermion propagator in HF approximation. Dashed line: free propagator,
solid line: dressed propagator.}
\end{figure}
The self-consistent mass can be thought of as the sum over all one-particle-irreducible (1PI) cactus type diagrams, see Fig.~2. 
\begin{figure}
\begin{center}
\epsfig{file=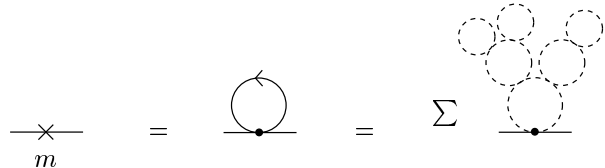,width=8.0cm,angle=0}
\end{center}
\caption{Dynamically generated fermion mass in the vacuum. A typical ``cactus" diagram produced by an iterative solution of
the HF equation in Fig.~1 is illustrated.}
\end{figure}
The (self-consistent) tadpole diagram receives contributions from all negative energy occupied states, leading to the gap equation
\begin{equation}
m\left(1-  \frac{Ng^2}{\pi} \ln \frac{\Lambda}{m}\right)=0
\label{p2}
\end{equation}
(with $\Lambda/2$ as UV cutoff \cite{C7}). Clearly, since the physical fermion mass is a pure manifestation of the Dirac sea,
it has to be put in by hand into an effective theory as an effective mass term
\begin{equation}
{\cal L}_{\rm eff}^{(1)}= - m \bar{\psi}\psi.
\label{p3}
\end{equation}   
(From now on the superscript on ${\cal L}_{\rm eff}$ refers to the number of loops of the self-energy diagram from which
it was derived. Due to resummations, it may contain higher loop effects as well.)

Let us now turn to the problem of finite fermion number. The HF approach still has the same basic structure as in Fig.~1,
except that the fermion self-energy is in general $x$-dependent. The propagator gets an extra contribution from the 
positive energy valence states. We denote the vacuum and valence particle contributions by ``$-$" and ``$+$", respectively.
For the free massive propagator for instance, the corresponding decomposition can be inferred from the well known
result for finite temperature and chemical potential \cite{C12,C12a},
\begin{eqnarray}
S(p)&=& S_-(p)+S_+(p),
\nonumber \\
S_-(p) &=& \frac{\rm i}{p\!\!\!/-m+{\rm i}\epsilon},
\nonumber \\
S_+(p) & = & -2\pi \delta(p^2-m^2)(p\!\!\!/+m)
\label{p4} \\
& \times & \left[\theta(-p_0)\theta(E_f-E_p)+\theta(p_0)
\theta(-E_f-E_p)\right].
\nonumber
\end{eqnarray}
Here, $S_-$ is just the free Feynman propagator, $E_f$ the Fermi energy or chemical potential.
The one-loop self-energy (i.e., the tadpole) then naturally splits up into the two pieces shown in Fig.~3.
\begin{figure}
\begin{center}
\epsfig{file=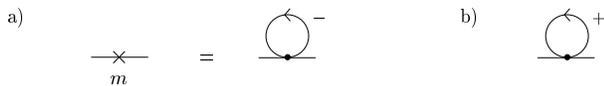,width=8.0cm,angle=0}
\end{center}
\caption{Decomposition of the fermion self-energy into a) negative and b) positive energy contributions.}
\end{figure}
In the effective theory with the Dirac sea integrated out, diagram a) is accounted for by the mass term (\ref{p3}).
Diagram b) corresponds to the lowest order ``no-sea" HF calculation with the original four-fermion
interaction from Eq.~(\ref{p1}).
This splitting of the tadpole into $+/-$ pieces is also the key for the following systematic procedure.
For a given number of loops, draw all topologically distinct 1PI cactus diagrams as shown in Fig.~4 up to five loops.
\begin{figure}
\begin{center}
\epsfig{file=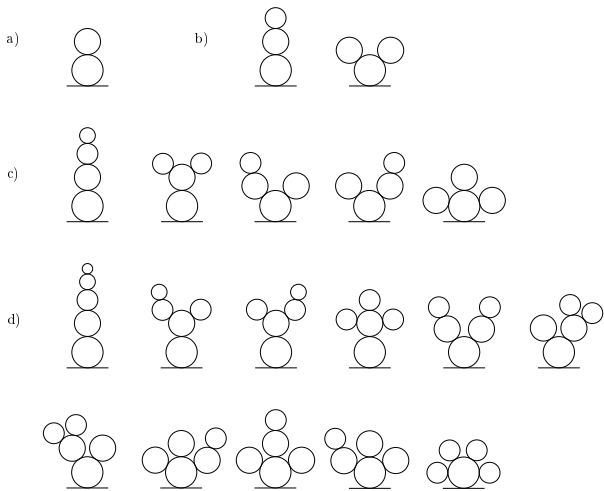,width=8.0cm,angle=0}
\end{center}
\caption{Topologically distinct diagrams contributing to the fermion self-energy in HF approximation
from two up to five loops. They are needed to construct the effective action for the GN$_2$ model in NNLO, see the main text.}
\end{figure}
Split each loop into a sum of $``+"$ and $``-"$ contributions. This generates $2^n$ distinct, ``labeled" diagrams out of each $n$-loop diagram
in Fig.~4. Then analyze each labeled diagram with respect to the question whether it is accounted for by the effective
action with the Dirac sea integrated out. If it is not, add a new term to ${\cal L}_{\rm eff}$ which generates the corresponding 
fermion self-energy.
 
Although this scheme is straightforward and constructive, it is still too naive. We will see that it would be meaningless
to terminate the procedure at any finite number
of loops. Remember that the mass term (\ref{p3}) already includes an infinite number of diagrams 
due to the self-consistency condition (see Fig.~2). Otherwise, dimensional transmutation, a truly
non-perturbative phenomenon, could not occur. We can only make sense out of the $n$-loop contribution to
the effective Lagrangian if we perform appropriate resummations. This turns out to be a crucial element of the whole
procedure and the way in which the bare coupling constant $g^2$ gets traded for a finite effective 
coupling constant.

To explain how this works, let us go through the first few terms of the loop expansion in some detail.
Guided by the DHN baryon at small filling, we shall treat $\bar{\psi}\psi$ and $k^2$ (or $\square$) as being of the same order
$\epsilon$ and derive systematically all terms up to O($\epsilon^4$).

At two-loop order, the only diagram (out of 4 labeled diagrams) which induces a new term in ${\cal L}_{\rm eff}$ is the
one shown in Fig.~5a. 
\begin{figure}
\begin{center}
\epsfig{file=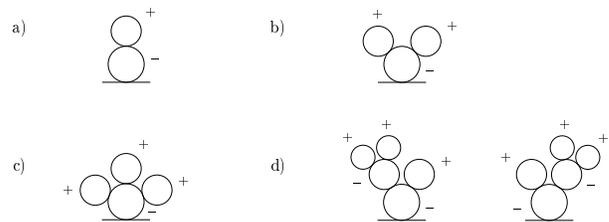,width=8.0cm,angle=0}
\end{center}
\caption{Labeled diagrams selected from the diagrams in Fig.~4, which are the seed for new terms in 
the effective Lagrangian. See main text for a detailed discussion.}
\end{figure}
This corresponds to a vertex correction to the $``+"$ tadpole in Fig.~3b. Technically, it involves the 
scalar vacuum polarization loop (labeled $``-"$ in Fig.~5a), where scalar means that the vertices are just 1. We
evaluate it by standard Feynman rules and expand the result in powers of the 4-momentum $k$ flowing 
through the graph. The result for the sum of the $``+"$ tadpole and Fig.~5a then gives the following self energy contribution,
\begin{eqnarray}
\delta \Sigma(k) & = &
- g^2 \langle \bar{\psi}\psi \rangle_k
\left\{1+ \frac{Ng^2}{\pi}\left( \ln \frac{\Lambda}{m}-1 \right. \right.
\nonumber \\
& & + \left. \left. \frac{k^2}{12 m^2} + \frac{(k^2)^2}{120 m^4} + {\rm O}(k^6)\right) \right\}.
\label{p5}
\end{eqnarray}
Here, the scalar condensate in momentum space, $\langle \bar{\psi}\psi \rangle_k$,
is defined so as to include only $``+"$ terms.
Whereas the $k^2$-dependent correction terms are suppressed (remember that $k^2 \sim \epsilon$),
the $k=0$ contribution $\sim \ln(\Lambda/m)$ needs a special treatment.
According to the gap equation (\ref{p2}), this term is actually of O(1) so that the series expansion
starting with Eq.~(\ref{p5}) is meaningless as it stands.
We therefore sum the whole series of graphs shown in Fig.~6
\begin{figure}
\begin{center}
\epsfig{file=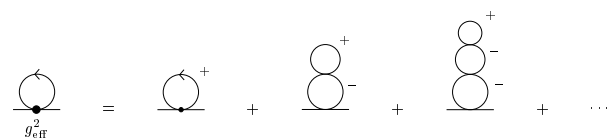,width=8.0cm,angle=0}
\end{center}
\caption{Resummation of scalar vacuum polarization graphs into a momentum dependent effective coupling. 
The infinite sum of $``-"$ bubbles may be thought of as generating the $\sigma$ propagator.}
\end{figure}
into a geometrical series, or, equivalently, replace the 
bare coupling constant by the effective, momentum dependent coupling
\begin{equation}
g_{\rm eff}^2(k)=
\frac{g^2}{1-  \frac{Ng^2}{\pi}\left( \ln \frac{\Lambda}{m}-1 +
\frac{k^2}{12 m^2} + \frac{(k^2)^2}{120 m^4} \right)}.
\label{p7}
\end{equation}
Due to the gap equation, the 1 in the denominator is now cancelled against $(Ng^2/\pi) \ln (\Lambda/m)$, and the bare
coupling constant $g^2$ drops out of this expression. Expanding the $k$-dependent terms again
[to O($k^4$)], we arrive at
the following momentum dependent effective coupling,
\begin{equation}
g_{\rm eff}^2 (k)  =  \frac{\pi}{N} \left( 1 + \frac{k^2}{12 m^2} + \frac{11 (k^2)^2}{720 m^4}\right) .
\label{p8}
\end{equation}
It is now easy to write down a two-loop effective Lagrangian which would give just these correction terms (correct
at two-loop level, but containing higher order terms due to resummation),
\begin{eqnarray}
{\cal L}_{\rm eff}^{(2)} &=& \frac{\pi}{2N} (\bar{\psi}\psi)^2 - \frac{\pi}{24 N m^2} (\square \bar{\psi}
\psi) (\bar{\psi}\psi) 
\nonumber \\
& & + \frac{11 \pi}{1440 N m^4}(\square^2 \bar{\psi}\psi) (\bar{\psi}\psi).
\label{p9}
\end{eqnarray}
Notice that the bare coupling constant $g^2$ in the original $(\bar{\psi}\psi)^2$
term has been replaced by $\pi/N$, and new four-fermion interaction terms containing derivatives 
of $\bar{\psi}\psi$ have been generated. Higher order terms could easily be obtained from the full $k$-dependence of the 
vacuum polarization graph if desired. The result is manifestly non-perturbative, as it does not contain $g^2$, and finite
owing to the use of the gap equation.

It may be worthwhile to pause here and comment on the value of the leading order effective coupling, $g_{\rm eff}^2=\pi/N$  [see Eq.~(\ref{p8})].
This quantity already appeared in the original paper by Gross and Neveu \cite{C1} as fermion-fermion scattering amplitude
at zero momentum. These authors also discuss the scalar $\sigma$ meson, pointing out that the square of the fermion-antifermion-$\sigma$
coupling constant is given by $g_{\sigma F \bar{F}}^2=4\pi m^2/N$. 
One can then understand the effective coupling in a similar way as in the old Fermi theory of weak interactions, namely as a product
of two coupling constants  and a heavy boson propagator ($M_{\sigma}=2m$),
\begin{equation}
g_{\sigma F \bar{F}} \frac{1}{M_{\sigma}^2} g_{\sigma F \bar{F}} = \frac{\pi}{N}.
\label{p9a}
\end{equation}
Keeping the momentum dependent terms approximately accounts for the finite range of the $\sigma$-exchange.

We now turn to the three-loop graphs. The two topologically distinct graphs in Fig.~4b give rise to $2\times2^3=16$ labeled subgraphs.
By inspection we find that only the graph shown in Fig.~5b is the seed for a new term in the effective action, all other
diagrams being generated by lower order terms or self-consistency in the $``+"$ sector. Evaluating the fermion loop with 
three scalar vertices and finite momenta $k_1,k_2$ from the two $``+"$ loops, we find the contribution to the self-energy
\begin{eqnarray}
\delta \Sigma(k) &=&
- \int \frac{{\rm d}k_1}{2\pi}\frac{{\rm d}k_2}{2\pi} (2\pi)\delta(k-k_1-k_2) 
\frac{(Ng^2)^3}{2\pi m}
\nonumber \\
& \times & \left( 1 + \frac{k_1^2+k_2^2+k_1k_2}{6m^2}\right) \frac{\langle \bar{\psi}\psi \rangle_{k_1}
\langle \bar{\psi}\psi \rangle_{k_2}}{N^2}.
\label{p10}
\end{eqnarray}
Here, it is sufficient to go to O($k^2$).
Resumming bubbles by replacing each of the three couplings $Ng^2$ at the vertices by effective ones
analogously to Fig.~6,
\begin{equation}
(Ng^2)^3 \to Ng_{\rm eff}^2(k_1)Ng_{\rm eff}^2(k_2)Ng_{\rm eff}^2(k_1+k_2),
\label{p11}
\end{equation}
we find to O($k^2$)
\begin{eqnarray}
\delta \Sigma(k) &=&
- \int \frac{{\rm d}k_1}{2\pi} \frac{{\rm d}k_2}{2\pi}(2\pi)\delta(k-k_1-k_2)
 \frac{\pi^2}{2m}
\label{p12} \\
& \times & \left( 1+ \frac{k_1^2+k_2^2+k_1k_2}{3m^2}\right) \frac{
\langle \bar{\psi}\psi \rangle_{k_1}\langle \bar{\psi}\psi \rangle_{k_2}}{N^2}.
\nonumber
\end{eqnarray}
The effective Lagrangian that yields this self-energy includes the following six-fermion interactions,
\begin{equation}
{\cal L}_{\rm eff}^{(3)}= \frac{\pi^2}{6mN^2}(\bar{\psi}\psi)^3 - \frac{\pi^2}{12 m^3 N^2}(\square \bar{\psi}\psi)
(\bar{\psi}\psi)^2.
\label{p13}
\end{equation}

At four-loop order, there is again only a single graph (out of $5\times2^4=80$ labeled diagrams, see Fig.~4c)
which generates a new term in the effective action. It is shown in Fig.~5c.
The central loop labeled $``-"$ with four scalar vertices is easy to compute, since we only need its value at $k=0$.
Resumming the bubbles by substituting $Ng^2\to \pi$ [to O($k^0$)], an eight-fermion interaction, 
\begin{equation}
{\cal L}_{\rm eff}^{(4)} = \frac{\pi^3}{24 m^2 N^3} (\bar{\psi}\psi)^4,
\label{p14}
\end{equation}
is induced.
This is not yet the whole story to O($\epsilon^4$) though. Out of the $11 \times 2^5=352$ labeled five-loop
diagrams derived from Fig.~4d, the two shown
in Fig.~5d give rise to a further eight-fermion interaction term of the same order as (\ref{p14}). Applying the by now
familiar resummation it is given by
\begin{equation}
{\cal L}_{\rm eff}^{(5)} = \frac{\pi^3}{8 m^2 N^3} (\bar{\psi}\psi)^4.
\label{p15}
\end{equation} 
The different origin of the contributions (\ref{p14}) and (\ref{p15}) becomes clearer if one interprets the scalar bubble sum as
$\sigma$-meson propagator. While the four-loop contribution (\ref{p14}) corresponds to a $\sigma \sigma \to \sigma \sigma$ point-like
interaction due to a heavy fermion loop, the five-loop term (\ref{p15}) has a different topology. It arises from the process
$\sigma \sigma \to \sigma \to \sigma \sigma$ with an iterated effective three-$\sigma$ vertex, see Fig.~7.
\begin{figure}
\begin{center}
\epsfig{file=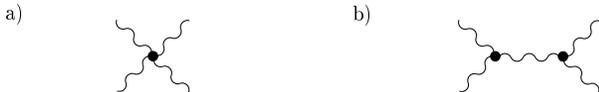,width=8.0cm,angle=0}
\end{center}
\caption{Illustration of the difference between the eight-fermion interactions ${\cal L}_{\rm eff}^{(4)}$ and ${\cal L}_{\rm eff}^{(5)}$
in Eqs.~(\ref{p14}) and (\ref{p15}). The wiggly lines are $\sigma$ mesons and represent the bubble sum shown in Fig.~6.
Each external line ends with a scalar density $\bar{\psi}\psi$.}
\end{figure}

This completes the discussion of all terms up to O($\epsilon^4$). We summarize by writing down the effective low-energy Lagrangian
for the massless GN$_2$ model valid to O($\epsilon^4$) where the Dirac sea has been integrated out,
\begin{eqnarray}
{\cal L}_{\rm eff} &=& \bar{\psi} ( {\rm i} \partial \!\!\!/  - m) \psi + \frac{\pi}{2N}(\bar{\psi}\psi)^2 
- \frac{\pi}{24 m^2 N} (\square \bar{\psi}\psi) (\bar{\psi}\psi)
\nonumber \\
&+ &    \frac{\pi^2}{6mN^2}(\bar{\psi}\psi)^3  
+  \frac{11 \pi}{1440 m^4 N}(\square^2 \bar{\psi}\psi) (\bar{\psi}\psi)
\nonumber \\
&- &  \frac{\pi^2}{12 m^3 N^2}(\square \bar{\psi}\psi) (\bar{\psi}\psi)^2
+  \frac{\pi^3}{6 m^2 N^3} (\bar{\psi}\psi)^4 .
\label{p16}
\end{eqnarray}
By counting powers of $\bar{\psi}\psi$ and $\square$, we identify 
the first two terms as leading order (LO), the next two as next-to-leading order (NLO) and the remaining three as
NNLO approximation.
To LO, a mass term is generated and the coupling constant $g^2$ of the four-fermion interaction 
is replaced by the effective coupling constant $\pi/N$. In higher orders, the Dirac sea manifests itself through  
momentum dependent couplings and the appearance of six- and eight-fermion interactions, to the
order we are working.
Here we are in the fortunate position of being able to evaluate all coefficients of the effective Lagrangian 
from the underlying field theory. Technically, the calculation of the coefficients only involves standard 
one-loop Feynman diagrams in the vacuum, without any reference to finite fermion density or baryons.
Nevertheless, the Lagrangian (\ref{p16}) should be adequate to predict properties of baryons or baryonic matter
by means of a purely classical calculation.

We shall test this conjecture against known results for the full GN$_2$ model in two different ways. First, we
do a ``no-sea" HF calculation
for matter at low density, assuming unbroken translational invariance. Second, we evaluate the baryon with small occupation
of the valence level by solving a non-linear Dirac equation.

If the condensate $\langle \bar{\psi}\psi\rangle$ is assumed to be translationally invariant, only the non-derivative terms
of the effective action enter, i.e.,
\begin{eqnarray}
{\cal L}_{\rm eff}' &=& \bar{\psi}({\rm i}\partial \!\!\!/ -m) \psi  +
\frac{\pi}{2N} (\bar{\psi}\psi)^2 + \frac{\pi^2}{6m N^2}(\bar{\psi}\psi)^3
\nonumber \\
& &  + \frac{\pi^3}{6m^2 N^3}(\bar{\psi}\psi)^4.
\label{p17}
\end{eqnarray}
The Euler-Lagrange equation allows us to identify the effective mass $M$ via
\begin{equation}
\frac{\partial}{\partial \bar{\psi}}{\cal L}_{\rm eff}' = ({\rm i}\partial \!\!\!/-M)\psi = 0
\label{p18} 
\end{equation}
with
\begin{equation}
M = m - \frac{\pi}{N} \langle \bar{\psi}\psi\rangle - \frac{\pi^2}{2m N^2}\langle \bar{\psi}\psi \rangle^2
- \frac{2\pi^3}{3 m^2 N^3}\langle \bar{\psi}\psi\rangle^3.
\label{p18a}
\end{equation}
By construction, the condensate $\langle \bar{\psi}\psi \rangle$ that appears here refers only to the positive energy sector. 
Using the matter part $S_+$ of the Dirac propagator (\ref{p4}) ($p_f=\sqrt{E_f^2-M^2}$ denotes the Fermi momentum),
we find
\begin{equation}
\langle \bar{\psi}\psi \rangle 
 =  N \frac{M}{\pi} \ln \left( \frac{p_f+\sqrt{p_f^2+M^2}}{M}\right). 
\label{p19}
\end{equation}
A low density expansion in $p_f$ yields 
\begin{equation}
M=m-p_f- \frac{p_f^2}{2m}- \frac{p_f^3}{2m^2} \pm ...
\label{p20}
\end{equation}
for the solution of Eqs.~(\ref{p18a},\ref{p19}),
in perfect agreement with the Taylor expansion of the exact result \cite{C2},
\begin{equation}
M=\sqrt{m^2-2m p_f}.
\label{p21}
\end{equation}
As is well known, this particular HF solution is not a minimum of the action. If translational invariance is 
assumed to be unbroken,
there is a first order phase transition which requires a Maxwell construction, see Ref.~\cite{C2}.
Nevertheless this unphysical solution can be used as a purely algebraic test of the effective action. We thus confirm
that the effects of the Dirac
sea are encoded in the effective action, both through the mass and coupling constant of the four-fermion
interaction and through induced many-fermion interaction terms.
Obviously, the lower the density one is interested in, the smaller the number of terms that 
need to be kept in ${\cal L}_{\rm eff}$.

A physically more relevant test case is the DHN baryon. We start with the Euler-Lagrange equation derived
from the full effective Lagrangian (\ref{p16}) and everywhere replace $\bar{\psi}\psi$ by its ground state
expectation value, this time determined by a single valence state.
Let us define the $x$-dependent condensate 
\begin{equation}
s=-\frac{\pi}{N} \langle \bar{\psi}\psi \rangle = - \pi \nu \bar{\psi}_0\psi_0
\label{p22}
\end{equation}
with $\nu=n/N$. Since $s(x)$ is time independent, we get
\begin{equation}
\square \langle \bar{\psi}\psi\rangle  =  \frac{N}{\pi} s''.
\label{p23}
\end{equation}
Consequently, the Euler-Lagrange equation can then be cast into the form of the following non-linear Dirac equation for the 
valence level,
\begin{equation}
\left( - \gamma_5 {\rm i} \partial_x + \gamma^0 S \right) \psi_0=E_0 \psi_0,
\label{p24}
\end{equation}
where the scalar potential $S$ is given by
\begin{eqnarray}
S & = & m+s+ \frac{1}{12m^2}s'' - \frac{1}{2m} s^2
+ \frac{11}{720m^4} s^{IV}
\nonumber \\
& & - \frac{1}{6m^3} \left[ 2 s'' s +(s')^2 \right]
+ \frac{2}{3m^2}s^3,
\label{p25}
\end{eqnarray}
and $s$ in turn depends on $\psi_0$ through Eq.~(\ref{p22}).
Eqs.~(\ref{p24},\ref{p25}) have to be solved self-consistently subject to the normalization condition
\begin{equation}
\int {\rm d}x \psi_0^{\dagger}\psi_0 = 1.
\label{p26}
\end{equation} 
The mass of the baryon in the effective theory is then calculated as follows: Deduce the 
Hamiltonian density from ${\cal L}_{\rm eff}$ (\ref{p16}) in the standard way,
\begin{equation}
{\cal H}_{\rm eff} = \frac{\partial {\cal L}_{\rm eff}}{\partial \dot{\psi}}\dot{\psi}-{\cal L}_{\rm eff}.
\label{p27}
\end{equation}
Since ${\cal L}_{\rm eff}$ is linear in $\dot{\psi}$, this amounts to dropping the term containing the 
time derivative of $\psi$ and reverting the overall sign of ${\cal L}_{\rm eff}$. The baryon mass is now simply given by
\begin{equation}
M_B=\int {\rm d}x {\cal H}_{\rm eff}.
\label{p28}
\end{equation}
To solve the ``no-sea" Dirac-HF equation is not easy and would have to be done numerically in
general. Since we know the exact answer for $\psi_0$ from the full GN$_2$ model and are primarily
interested in checking our effective action, we proceed differently. 
We use the following representation of the $\gamma$ matrices,
\begin{equation}
\gamma^0 = - \sigma_1, \quad \gamma^1 = {\rm i}\sigma_3, \quad \gamma_5=\gamma^0 \gamma^1 = - \sigma_2.
\label{p1a}
\end{equation}
The (positive energy) valence
level of the DHN baryon then has the spinor wave function and energy \cite{C7}
\begin{equation}
\psi_0(x) =\frac{\sqrt{ym}}{2} 
\left(
\begin{array}{r} \frac{1}{\cosh \xi_-}\\ - \frac{1}{\cosh \xi_+}\end{array}
\right),
\quad
E_0 = m \sqrt{1-y^2},
\label{p29}
\end{equation}
with the definitions
\begin{equation}
\xi_{\pm} = ymx \pm \frac{1}{2} {\rm artanh}\, y,
\quad
y = \sin \theta, \quad \theta= \frac{\pi \nu}{2}.
\label{p30}
\end{equation}
We plug this ansatz into the non-linear Dirac equation (\ref{p24}), expand in the filling fraction $\nu$ and check with MAPLE that 
the equation is indeed satisfied up to (including) O($\nu^6$). For the baryon mass (\ref{p28}) we obtain at this order
\begin{equation}
M_B= nm \left( 1-\frac{\pi^2\nu^2}{24} + \frac{\pi^4 \nu^4}{1920}-\frac{\pi^6\nu^6}{322560}\right).
\label{p31}
\end{equation}
We have truncated the series because higher order terms are not reliable, as they would require an improved
effective action. The exact DHN baryon mass is given by the compact formula
\begin{equation}
M_B= n m \frac{\sin \theta}{\theta} 
\label{p32}
\end{equation}
with $\theta$ as defined in Eq.~(\ref{p30}).
If we expand this function in powers of $\nu$, the series indeed starts with (\ref{p31}). This is obviously a very good test
of our effective action and shows that the terms retained as well as the resummations done are consistent.   

In the massless GN$_2$ model, the only regime where an almost local effective action for the baryon can be justified
is $\nu\ll 1$, i.e., small valence filling fraction. Once we include a bare mass term, there is yet another handle to suppress
Dirac sea effects and make the scalar potential softer, namely by increasing the bare mass. This incites us to
generalize the ``no-sea" effective action to the massive GN$_2$ model.
The only difference to the previous calculation comes from the modified gap equation which now reads \cite{C13}
\begin{equation}
\frac{\pi}{Ng^2} = \gamma + \ln \frac{\Lambda}{m},  \quad \gamma:=\frac{\pi}{Ng^2}\frac{m_0}{m}.
\label{p33}
\end{equation}
While the diagrams used to derive the effective action are the same as above, the algebra slightly changes. 
We immediately jump to the final effective Lagrangian for the massive GN$_2$ model in NNLO,
\begin{eqnarray}
{\cal L}_{\rm eff} &=& \bar{\psi}\left( {\rm i}\partial \!\!\!/ -m\right) \psi  + \frac{\pi}{2N} \frac{1}{(1+\gamma)}(\bar{\psi}\psi)^2
\nonumber \\
& -& \frac{\pi}{24 m^2 N}\frac{1}{(1+\gamma)^2}(\square \bar{\psi}\psi)(\bar{\psi}\psi)
\nonumber \\
&+ &  \frac{\pi^2}{6mN^2}\frac{1}{(1+\gamma)^3} (\bar{\psi}\psi)^3 
\nonumber  \\
& + & \frac{11\pi}{1440 m^4 N}\frac{(1+6\gamma/11)}{(1+\gamma)^3} (\square^2\bar{\psi} \psi)
(\bar{\psi}\psi)
\nonumber \\
&- &  \frac{\pi^2}{12 m^3 N^2}  \frac{(1+\gamma/2)}{(1+\gamma)^4} (\square \bar{\psi}\psi)(\bar{\psi}\psi)^2
\nonumber \\
& + &  \frac{\pi^3}{6 m^2 N^3} \frac{(1+\gamma/4)}{(1+\gamma)^5}(\bar{\psi}\psi)^4 .
\label{p35}
\end{eqnarray}
It reduces to Eq.~(\ref{p16}) in the chiral limit or, equivalently, at $\gamma=0$. 
In order to test our effective Lagrangian, we turn to the 
known exact baryons of the massive GN$_2$ model \cite{C14,C15,C15a}. Defining $s(x)$
as in Eq.~(\ref{p22}), the non-linear
Dirac equation (\ref{p24}) now contains the scalar potential
\begin{eqnarray}
S &=& m+ \frac{1}{(1+\gamma)}s + \frac{1}{12m^2}\frac{1}{(1+\gamma)^2}s''
\nonumber \\
& - &  \frac{1}{2m} \frac{1}{(1+\gamma)^3}s^2 
+   \frac{11}{720m^4} \frac{(1+6\gamma/11)}{(1+\gamma)^3}s^{IV}
\label{p36} \\
& - & \frac{1}{6m^3}\frac{(1+\gamma/2)}{(1+\gamma)^4}\left[2 s'' s
+(s')^2 \right]+ \frac{2}{3m^2} \frac{(1+\gamma/4)}{(1+\gamma)^5}s^3.
\nonumber
\end{eqnarray}
The exact spinor wave function and the energy remain the same as 
in Eqs.~(\ref{p29}), but the relationship between the parameter $y$ and the valence occupation fraction $\nu$
changes to
\begin{equation}
\frac{\nu}{2}=\frac{\theta}{\pi}+ \frac{\gamma}{\pi}\tan \theta, \quad y=\sin \theta.
\label{p37}
\end{equation}
Eq.~(\ref{p37}) can be solved for $\theta$ by means of a power series expansion in $\nu$ for given $\gamma$.
The exact baryon mass is given by
\begin{equation}
M_B=\frac{2mN}{\pi}\left[\sin \theta + \gamma\, {\rm artanh}\,(\sin \theta) \right].
\label{p38}
\end{equation}
We have verified with MAPLE that the full wave function $\psi_0$ solves the non-linear Dirac equation
derived from Eq.~(\ref{p35}) to an accuracy of O($\nu^6$), as in the massless case. The baryon mass calculated from 
the effective action is found analytically to be
\begin{eqnarray}
M_B & = & nm \left( 1- \frac{\pi^2\nu^2}{24(1+\gamma)^2}+\frac{\pi^4(1+9\gamma)\nu^4}{1920(1+\gamma)^5}\right.
\nonumber \\
& & - \left. \frac{\pi^6(1-54\gamma+225\gamma^2)\nu^6}{322560(1+\gamma)^8}\right),
\label{p39}
\end{eqnarray}
the generalization of Eq.~(\ref{p31}) to arbitrary $\gamma$. To the given order this agrees once again with the
result based upon the series expansion of the exact equations (\ref{p37},\ref{p38}).

These results show that with our ``no-sea" effective action, we have indeed achieved what we were aiming at.
Depending on the number of terms 
one is willing to include, one can systematically get an increasingly accurate value for the baryon mass
and valence wave function. Note that the expansion parameter changes from $\nu$ in the chiral limit to $\nu / \gamma$
in the heavy quark limit. As a matter of fact, the truncated series (\ref{p39}) is a much better guide to the exact
baryon mass in the full range of the parameters ($\nu,\gamma$) than could have been expected on the basis of
our derivation. We find that the NNLO effective action yields the binding energy 
of the baryon with an error of less than 1.5$\%$ even for full occupation ($\nu=1$) and arbitrary $\gamma$.
In the chiral limit, the error is below 0.3$\%$ for all values of $\nu$.
Even the LO effective action yields results no worse than 10-20$\%$ for the binding energy. 

The most important result of this section is the effective Lagrangian (\ref{p35}) for the massive GN$_2$ model.
It contains the corresponding Lagrangian (\ref{p16}) for the massless GN$_2$ model as a special case ($\gamma=0$).
The applications to homogeneous matter and baryons show that our general scheme is correct
and underline the systematic character of the expansion.
In the derivation, we only had to compute standard one-loop Feynman diagrams with increasing number of external lines.
Nevertheless, we demonstrated that an excellent approximation to the baryon mass can be obtained even
in regions of the parameter space where we had no a-priori reason to assume that our truncation of the
effective action was justified.
\section{Nambu--Jona-Lasinio model (NJL$_2$)}
The Lagrangian of the Gross-Neveu model with continuous chiral symmetry (NJL$_2$ model) is
\begin{equation}
{\cal L}=  \bar{\psi} {\rm i} \partial \!\!\!/ \psi + \frac{g^2}{2} \left[ (\bar{\psi}\psi)^2 + (\bar{\psi}{\rm i}\gamma_5 \psi)^2\right].
\label{p40}
\end{equation}
In this model, the Dirac sea has a more dramatic impact on the hadrons of the theory than in the case of the GN$_2$ model.
This is already clear from the existence of massless mesons and baryons, despite the fact that the elementary fermions 
acquire a dynamical mass \cite{C2}. As compared to the GN model, one has to expect additional complications
when integrating out the Dirac sea. 
In the present section, we therefore restrict ourselves to the LO calculation of the effective action.
It turns out that the LO calculation in the NJL$_2$ model is about as complex as the NNLO calculation 
in the GN$_2$ model described in Sect.~II.

The general procedure and the diagrams to be considered stay the same as before, except that each vertex
can now be either 1 (scalar) or ${\rm i}\gamma_5$ (pseudoscalar). Some diagrams vanish due to parity selection rules.

At one-loop order (tadpole), the scalar contribution yields an effective mass just like in the GN$_2$ model. The pseudoscalar
tadpole vanishes, so that the one-loop contribution to ${\cal L}_{\rm eff}$ is again a standard mass term,
\begin{equation}
{\cal L}_{\rm eff}^{(1)}= - m \bar{\psi}\psi.
\label{p40a}
\end{equation}
At two-loop level, we have to consider that the two vertices in Fig.~5a can both be either scalar or pseudoscalar. The first case
is identical to the GN$_2$ model and, after resumming the scalar bubbles,  yields the first term in Eq.~(\ref{p9})
(higher order terms are discarded since we now work to LO only). The second case 
involves a pseudoscalar vacuum polarization. For the sum of the $``+"$ tadpole and this correction,
we obtain to O($k^2$)
\begin{equation}
\delta \Sigma(k)=-g^2 \langle \bar{\psi}{\rm i}\gamma_5\psi \rangle_k
\left\{1+\frac{Ng^2}{\pi}\left(\ln \frac{\Lambda}{m} + \frac{k^2}{4m^2}\right)\right\}.
\label{p42}
\end{equation}
This should be compared to the corresponding scalar result, Eq.~(\ref{p5}). Using the same arguments for resumming 
the inner ``$-$" bubbles as in the scalar case, we get (denoting the pseudoscalar coupling by $G_{\rm eff}$)
\begin{equation}
G_{\rm eff}^2(k) =
\frac{g^2}{1-  \frac{Ng^2}{\pi}\left( \ln \frac{\Lambda}{m} +
\frac{k^2}{4 m^2} \right)}.
\label{p44}
\end{equation}
Invoking the gap equation (\ref{p2}) now yields the result
\begin{equation}
G_{\rm eff}^2(k) = -\frac{4m^2}{k^2}\frac{\pi}{N}.
\label{p45}
\end{equation}
The pseudoscalar bubble sum has produced the pole of the massless ``pion". Actually, the difference between
(\ref{p45}) and the leading order result for the scalar coupling
$g_{\rm eff}^2=\pi/N$ can easily be understood: As already noted
by Gross and Neveu \cite{C1}, the coupling constants $g_{\pi F\bar{F}}$ and $g_{\sigma F\bar{F}}$ are identical
due to chiral symmetry. In Eq.~(\ref{p9a}), we have interpreted the scalar effective coupling constant in terms of
a $\sigma$-exchange, replacing the propagator by a constant, namely its value at $k^2=0$. 
However, this cannot be done in the case of the $\pi$-exchange. The limit $k^2\to0$ is singular and we must keep
the leading momentum dependence,
\begin{equation}
g_{\pi F\bar{F}}\frac{1}{-k^2}g_{\pi F \bar{F}}= - \frac{4m^2}{k^2}\frac{\pi}{N},
\label{p45a}
\end{equation} 
thus reproducing Eq.~(\ref{p45}).
Clearly, the pole in the effective pseudoscalar interaction has drastic consequences for the structure of the effective
theory. Since the expansion around $k=0$ is singular
in the NJL$_2$ model, we have to reconsider the ordering principle 
behind our approach. 
Keeping the momentum dependence from the pion pole in Eq.~(\ref{p45}),
the total two-loop contribution to the effective Lagrangian assumes the non-local form
\begin{equation}
{\cal L}_{\rm eff}^{(2)} = \frac{\pi}{2N}(\bar{\psi}\psi)^2 + \frac{2\pi m^2}{N} \bar{\psi}{\rm i}\gamma_5 \psi \frac{1}
{\square}\bar{\psi}{\rm i}\gamma_5 \psi.
\label{p46}
\end{equation}
As before we shall treat $\bar{\psi}\psi$ and $\square$ as being of O($\epsilon$). We would like to 
test our effective action against the multi-fermion bound states of Shei,  in analogy to our test 
against the DHN baryon in the GN$_2$ case.
From these known bound states we infer that $\bar{\psi}{\rm i}\gamma_5 \psi$ is of O($\epsilon^{3/2}$).
Hence both terms in Eq.~(\ref{p46}) are of the same O($\epsilon^2$), which defines our LO approximation.

At three-loop level we must again consider the graph in Fig.~5b. If all three couplings are scalar, it is of O($\epsilon^3$)
and therefore NLO. We neglect it in the present case. The only other 
non-vanishing contribution involves one scalar and two pseudoscalar couplings. We evaluate the corresponding 
Feynman graph and resum the scalar and pseudoscalar bubbles into effective couplings, generating
two pion poles. The resulting effective Lagrangian reads
\begin{equation}
{\cal L}_{\rm eff}^{(3)} = \frac{8\pi^2 m^3}{N^2}\bar{\psi}{\rm i}\gamma_5 \psi \frac{1}{\square}\bar{\psi}\psi \frac{1}
{\square}\bar{\psi}{\rm i}\gamma_5 \psi.
\label{p47}
\end{equation}
Notice that this is again of O($\epsilon^2$). Due to the inverse powers of momenta or $\square$, 
unlike in the GN$_2$ case, there is no suppression of the three-loop contribution
as compared to the two-loop contribution by a power of $\epsilon$.
Even worse, one can identify a whole class of higher terms with arbitrary many loops contributing to the 
same order of $\epsilon$.
This calls once again for a resummation, namely [add up the 2nd term in Eq.~(\ref{p46}) and Eq.~(\ref{p47})] 
\begin{eqnarray}
& & \frac{2\pi m^2}{N} \bar{\psi}{\rm i}\gamma_5 \psi \frac{1}{\square}
\left( 1+ \frac{4\pi m}{N} \bar{\psi}\psi \frac{1}{\square} + ...\right)
\bar{\psi}{\rm i}\gamma_5 \psi
\nonumber \\
&  &  \to \frac{2\pi m^2}{N} \bar{\psi}{\rm i}\gamma_5 \psi \frac{1}{\square-\frac{4\pi m}{N}\bar{\psi}\psi}\bar{\psi}{\rm i}\gamma_5 \psi .
\label{p48}
\end{eqnarray}
We have summed up a special class of higher order diagrams, singled out by being of O($\epsilon^2$). They
can be described as follows: Draw a string of pseudoscalar bubbles between two $\bar{\psi}{\rm i}\gamma_5\psi$
(valence) condensates. Then attach at most one scalar tadpole to any of the pseudoscalar loops (on
either side). All diagrams generated with this prescription are summed up.
If there are $n$ scalar condensates, this must be matched by $n$+1 pion poles.
It is easy to convince oneself that any other way of attaching tadpoles to the bubble graphs is punished by
a higher order in $\epsilon$. As a result of this discussion, we replace the effective action (\ref{p46}) induced by
the two-loop graph with
\begin{equation}
{\cal L}_{\rm eff}^{(2)}=  \frac{\pi}{2N}(\bar{\psi}\psi)^2 + 
\frac{2\pi m^2}{N} \bar{\psi}{\rm i}\gamma_5 \psi \frac{1}{\square-\frac{4\pi m}{N}
\bar{\psi}\psi}\bar{\psi}{\rm i}\gamma_5 \psi .
\label{p49}
\end{equation} 
The massless pion propagator has been superseded by the propagator of the pion in a scalar background
potential and sums up higher loop contributions of O($\epsilon^2$) in ${\cal L}_{\rm eff}$ as shown schematically in Fig.~8. 
\begin{figure}
\begin{center}
\epsfig{file=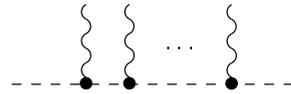,width=4.0cm,angle=0}
\end{center}
\caption{Dressed pion propagator (dashed line), connected to an arbitrary number
of scalar densities $\bar{\psi}\psi$ via $\sigma$-exchanges (wiggly lines). See Eqs.~(\ref{p48}) and (\ref{p49}).}
\end{figure}
Notice that the pion 
self-energy appearing in the denominator of Eq.~(\ref{p49}) has a form reminiscent of the Gell-Mann, Oakes,
Renner (GOR) relation \cite{C16} which relates the pion mass to the bare fermion mass and the chiral
condensate,
\begin{equation}
m_{\pi}^2 = -\frac{4\pi m_0}{N}\langle \bar{\psi}\psi \rangle.
\label{p50}
\end{equation}
However, in Eq.~(\ref{p49}), $m$ is the physical fermion mass, and $\bar{\psi}\psi$ refers to the ($x$-dependent) valence contribution 
to the condensate only. Nevertheless, the formal similarity is striking and may point to a deeper physics reason
behind our rather technical resummation. 

Due to the troublesome occurrence of inverse powers of $\epsilon$ in the pseudoscalar effective coupling, we have to go to even
higher order in the loop expansion in order to identify all LO terms. At four-loop order, we had to consider
Fig.~5c for the GN$_2$ model (where it was NNLO). If all four vertices are scalar (like in the GN$_2$ model), it is indeed of
O($\epsilon^4$) and can
be discarded for our present purpose. If two vertices are scalar and two are pseudoscalar, after resummation
it will be of O($\epsilon^3$) and hence still negligible. The only O($\epsilon^2$) contribution is the diagram in Fig.~5c
with four ${\rm i}\gamma_5$ vertices. The four pion pole terms together with four pseudoscalar condensates conspire
to give O($\epsilon^2$). The Feynman diagram calculation yields the effective action
\begin{equation}
{\cal L}_{\rm eff}^{(4)}= - \frac{32\pi^3m^6}{N^3} \left(\frac{1}{\square}\bar{\psi}{\rm i}\gamma_5 \psi
\right)^4.
\label{p50a}
\end{equation}
According to our previous treatment of the two-loop diagram, we should once again replace every denominator $\square$
by $\square-\frac{4\pi m}{N}\bar{\psi}\psi$, thereby summing up another infinite set of diagrams of O($\epsilon^2$)
but containing multi-fermion interactions of arbitrary order,
\begin{equation}
{\cal L}_{\rm eff}^{(4)}= - \frac{32\pi^3m^6}{N^3} \left(\frac{1}{\square-\frac{4\pi m}{N}\bar{\psi}\psi}
\bar{\psi}{\rm i}\gamma_5 \psi 
\right)^4.
\label{p51}
\end{equation}

Finally we turn to the five-loop graph in Fig.~5d, which contributed in NNLO in the GN$_2$ case. The interesting 
case is the one where there are four pseudoscalar vertices and one scalar vertex, the latter connecting the two $``-"$ bubbles. 
The calculation
including all necessary resummations yields
\begin{equation}
{\cal L}_{\rm eff}^{(5)}=\frac{32\pi^3m^6}{N^3} \left(\frac{1}{\square-\frac{4\pi m}{N}\bar{\psi}\psi}\bar{\psi}{\rm i}\gamma_5 \psi
\right)^4.
\label{p52}
\end{equation}
It yields exactly the same magnitude as (\ref{p51}) but the opposite sign, so that the four- and five-loop terms cancel in the NJL$_2$ model. 
In our opinion this cancellation is an expression of the low energy theorem according to which the $\pi \pi$ interaction at 
zero momentum should vanish \cite{C17}. Here this comes about as a result of a cancellation between the $4\pi$ vertex
and the process $\pi \pi \to \sigma \to \pi \pi$ involving the $\pi\pi\sigma$ vertex, see Fig.~9.
\begin{figure}
\begin{center}
\epsfig{file=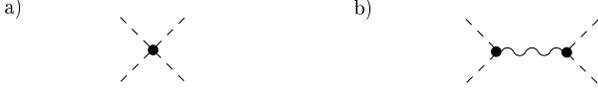,width=8.0cm,angle=0}
\end{center}
\caption{a) Effective direct 4$\pi$ interaction corresponding to Eq.~(\ref{p51}). b) Effective 4$\pi$ interaction via $\sigma$-exchange 
leading to Eq.~(\ref{p52}). The two diagrams cancel exactly due to the vanishing of the $\pi\pi$ interaction at zero momentum,
in accordance with low energy theorems.}
\end{figure}
It is interesting that the
NJL$_2$ model respects the low energy theorems in the large $N$ limit, in spite of the usual strong reservations against
Goldstone bosons in two dimensions \cite{C18}.

We have now identified all terms of O($\epsilon^2$) in the effective Lagrangian. As announced, the calculation was 
more involved than in the GN$_2$ model. Whereas
in the GN$_2$ case we had to deal with four-, six- and eight-fermion interactions in LO, NLO and NNLO, 
here we had to sum up terms corresponding to 2$n$-fermion interactions with arbitrary $n$ already in LO. The reason is the
massless pion pole which induces inverse powers of $\epsilon$. It is responsible for the non-locality
of the LO effective interaction for the NJL$_2$ model which nevertheless has a rather simple final form,
\begin{eqnarray}
{\cal L}_{\rm eff} &=&  \bar{\psi}({\rm i}\partial \!\!\!/-m)\psi+ \frac{\pi}{2N}(\bar{\psi}\psi)^2
\nonumber \\
& + & \frac{2\pi m^2}{N}\bar{\psi}{\rm i}\gamma_5 \psi \frac{1}{\square - \frac{4\pi m}{N}\bar{\psi}\psi}\bar{\psi}{\rm i}\gamma_5\psi.
\label{p53}
\end{eqnarray}
The pseudoscalar term is associated with the fermion-fermion interaction through $\pi$-exchange, see Fig.~10a. The pion
propagator has a long range and is dressed as in Fig.~8. The scalar term arises from $\sigma$-exchange, Fig.~10b, and
is of zero range in LO. It is common to the GN$_2$ and NJL$_2$ models. 
\begin{figure}
\begin{center}
\epsfig{file=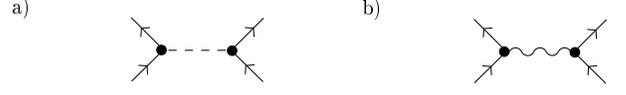,width=8.0cm,angle=0}
\end{center}
\caption{a) Effective fermion-fermion interaction through dressed $\pi$-exchange, giving rise to the non-local
term in ${\cal L}_{\rm eff}$, Eq.~(\ref{p53}). b) Effective fermion-fermion interaction through $\sigma$-exchange, yielding
the local $(\bar{\psi}\psi)^2$-term in ${\cal L}_{\rm eff}$.}
\end{figure}

At first glance, the non-locality of ${\cal L}_{\rm eff}$ is irritating, since
it is expected to make the solution of the Euler-Lagrange equation much harder. However in the present case, the
particular structure of the non-locality allows us to trade the non-local effective Lagrangian for a local 
one, at the cost of introducing an elementary pseudoscalar field $\Pi(x)$.
The local effective Lagrangian equivalent to (\ref{p53}) is given by
\begin{eqnarray}
{\cal L}_{\rm eff}^{\rm loc} &=&  \bar{\psi}({\rm i}\partial \!\!\!/-m)\psi+ \frac{\pi}{2N}(\bar{\psi}\psi)^2+ 
\frac{1}{2} \partial_{\mu}\Pi \partial^{\mu} \Pi
\nonumber \\
& + & \sqrt{\frac{4\pi}{N}}m \bar{\psi}{\rm i}\gamma_5 \psi \Pi
+ \frac{2\pi m}{N}\bar{\psi}\psi \Pi^2.
\label{p54}
\end{eqnarray}
Since we are in the large $N$ limit, $\Pi$ can be treated as a classical field. The equation of motion for
$\Pi$ following from (\ref{p54}) is
\begin{equation}
\left( \square - \frac{4\pi m}{N}\bar{\psi}\psi \right) \Pi = \sqrt{\frac{4\pi}{N}} m \bar{\psi}{\rm i}\gamma_5 \psi.
\label{p55}
\end{equation} 
Upon solving this inhomogeneous equation formally and plugging the result into Eq.~(\ref{p54}) we
indeed recover the non-local effective Lagrangian (\ref{p53}). So we wind up with
a local field theory containing both fermions and an elementary $\pi$-meson, the latter 
described by the field $\Pi$.

Since the derivation was quite complicated (in particular identifying all LO terms), it is again crucial
to test ${\cal L}_{\rm eff}$ against exact results for the NJL$_2$ model. For this purpose, we choose the multi-fermion
bound states of Shei \cite{C6}.
They are analogous to the DHN baryon of the GN$_2$ model in the way they were derived (inverse scattering theory), but in our
opinion carry vanishing baryon number and should be regarded as ``baryonium" states \cite{C19,C8}. 

Our strategy is as follows. Take the valence spinor wave function of Shei in the notation of Ref.~\cite{C8},
\begin{equation}
\psi_0=\sqrt{\frac{m|\cos \theta|}{2}}\frac{1}{\cosh \xi}\left( \begin{array}{c} \cos \theta/2 \\ \sin \theta/2 \end{array}\right)
\label{p56}
\end{equation}
with
\begin{eqnarray}
\xi &=& m x \cos \theta,
\nonumber \\
\theta &=& \left( \frac{3}{2}-\nu \right)\pi
\label{p57}
\end{eqnarray}
and energy $E_0=-m\sin \theta$. This yields the (valence) condensates
\begin{eqnarray}
\langle \bar{\psi}\psi \rangle & = & N m \nu \frac{\sin \pi \nu \cos \pi \nu}{2 \cosh^2 \xi},
\nonumber \\
\langle \bar{\psi}{\rm i}\gamma_5 \psi \rangle & = & N m \nu \frac{\sin^2 \pi \nu}{2 \cosh^2 \xi}.
\label{p58}
\end{eqnarray}
Eq.~(\ref{p55}) for $\Pi$  becomes 
\begin{equation}
\left( \partial_{\xi}^2+ \frac{2\pi \nu \cot \pi \nu}{\cosh^2 \xi}\right) \Pi = - \frac{\nu\sqrt{\pi N}}{\cosh^2\xi}
\label{p59}
\end{equation}
and, to leading order in $\nu$, is solved by
\begin{equation}
\Pi = - \frac{\nu \sqrt{\pi N}}{1+{\rm e}^{2\xi}}.
\label{p60}
\end{equation}
The Euler-Lagrange equation for $\psi_0$ is equivalent to the ``no-sea" Dirac-HF equation,
\begin{equation}
\left( \gamma_5 \frac{1}{\rm i} \partial_x + \gamma^0 S(x)+{\rm i}\gamma^1 P(x)\right)\psi_0=E_0\psi_0,
\label{p61}
\end{equation}
with scalar and pseudoscalar potentials
\begin{eqnarray}
S(x)&=& m - \frac{\pi}{N} \langle \bar{\psi}\psi \rangle - \frac{2\pi m}{N}\Pi^2
\ = \ m\left(1-\frac{2 \pi^2 \nu^2}{1+{\rm e}^{2\xi}}\right),
\nonumber \\
P(x) &=& - \sqrt{\frac{4\pi}{N}}m\Pi
\ = \ m \frac{2\pi \nu}{1+{\rm e}^{2\xi}}.
\label{p62}
\end{eqnarray}
We have kept the LO terms only. We find that $\psi_0$ satisfies the HF equation up to corrections of O($\nu^3$).
The Hamiltonian density is given by
\begin{eqnarray}
{\cal H} & = & \psi^{\dagger} \left( - \gamma_5 {\rm i}\partial_x + \gamma^0 m \right)\psi 
- \frac{\pi}{2N} (\bar{\psi}\psi)^2 + \frac{1}{2} (\partial_x \Pi)^2
\nonumber \\
& & - \sqrt{\frac{4\pi}{N}}m \bar{\psi}{\rm i}\gamma_5 \psi \Pi - \frac{2\pi m}{N}\bar{\psi}\psi \Pi^2.
\label{p63}
\end{eqnarray}
Taking its expectation value in the valence state, we insert the above results for $\psi_0$ and $\Pi$ and integrate over $x$.
The derivative term gives zero, the mass term has to be treated in NLO in $\nu$ and yields $N m\nu-Nm\pi^2\nu^3/2$, the four
remaining terms give $Nm\pi^2\nu^3/3$ to LO.
The total result for the mass adds up to
\begin{equation}
M=\int {\rm d}x {\cal H} = Nm\nu \left(1  - \frac{1}{6}\pi^2\nu^2\right),
\label{p64}
\end{equation}
in agreement with the first two terms of the Taylor expansion in $\nu$ of the exact result,
\begin{equation}
M=\frac{Nm}{\pi}\sin \pi \nu.
\label{p65}
\end{equation}
Since we cannot expect more from a LO calculation, the test was successful.

Now consider the question of baryon number. One interesting aspect of the NJL$_2$ model
is the fact that a topologically non-trivial mean field can induce fermion number in the Dirac sea
\cite{C20,C21}. Since this is a physical effect but we have eliminated the Dirac sea, it must show up somewhere
else in the effective theory. To understand what is happening,
we treat the fermion density $\psi^{\dagger}\psi=\bar{\psi}\gamma^0 \psi$ as an effective operator, in analogy with the 
scalar or pseudoscalar densities above. To this end we write down the dressed tadpole with a $\gamma^0$ vertex
and again resolve it into positive and negative energy contributions. The diagrams involving only
$``-"$ terms add up to the divergent density of the (vacuum) Dirac sea and can be subtracted.
The $``+"$ tadpole represents the density contribution of the valence level, $\rho_{\rm val}=n \psi_0^{\dagger}\psi_0$, see Fig.~11a.
Its first correction is driven by the two-loop graphs in Fig.~11. 
\begin{figure}
\begin{center}
\epsfig{file=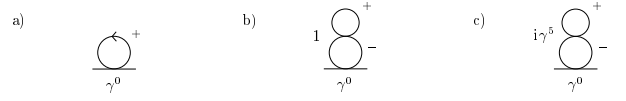,width=8.0cm,angle=0}
\end{center}
\caption{Illustration of the effective fermion density, Eq.~(\ref{p67}). a) Valence contribution. b) Vanishing sea contribution 
due to the scalar interaction. c) Induced fermion density due to the pseudoscalar
interaction.}
\end{figure}
If the upper vertex is 1 (scalar $``+"$ loop, Fig.~11b), the vacuum polarization bubble is the time component
of a four-vector and therefore proportional to $k^0$ (where $k$ is the external momentum). Since the self-energy
insertion is static, this term vanishes. This is the reason why we did not have to worry about 
induced fermion number in the GN$_2$ model. If the upper vertex is ${\rm i}\gamma_5$ (pseudoscalar
$``+"$ loop, Fig.~11c), the $``-"$ bubble behaves like the space component of a four-vector and is proportional to $k^1$.
We can replace the pseudoscalar $``+"$ bubble by the full pseudoscalar self-energy $P$, thereby
summing up all the LO diagrams as explained above. The result for the induced fermion density
in coordinate space then assumes the simple form
$\rho_{\rm ind} = \sqrt{N/\pi}\partial_x \Pi$,
where the derivative reflects the above mentioned $k^1$-dependence.
The total fermion density in the effective theory consists of the valence fermion density and the
induced one,
\begin{equation}
\rho = \rho_{\rm val}+\rho_{\rm ind}= n \psi_0^{\dagger}\psi_0 + \sqrt{\frac{N}{\pi}}\partial_x \Pi.
\label{p67}
\end{equation}
This allows us to correctly compute the fermion density without Dirac sea. If we insert the expressions
for the Shei bound state, we get an exact cancellation between the two contributions, in agreement
with two independent recent results \cite{C19,C8}. 

Let us now switch on the bare fermion mass in the NJL$_2$ model. As discussed above, this modifies the vacuum gap equation,
see Eq.~(\ref{p33}), and thereby the effective couplings. For the scalar coupling the corresponding
LO result can be read off from the $(\bar{\psi}\psi)^2$ term in Eq.~(\ref{p35}),
\begin{equation}
g_{\rm eff}^2 = \frac{\pi}{N}\frac{1}{(1+\gamma)}.
\label{p67a}
\end{equation} 
An analogous calculation for the pseudoscalar effective coupling yields
\begin{equation}
G_{\rm eff}^2 =\frac{4m^2}{4m^2\gamma-k^2} \frac{\pi}{N}.
\label{p68}
\end{equation}
The massless pion pole gets replaced by a massive one with the pion mass 
\begin{equation}
m_{\pi}^2=4m^2\gamma,
\label{p69}
\end{equation} 
in agreement with the leading order pion mass prediction near the chiral limit (GOR relation \cite{C2,C16}).
The next steps depend on the regime one is interested in. Two cases are particularly simple and 
will be considered here: i) $m_\pi^2$ and $k^2$ are both of O($\epsilon$), and the 
scalar and pseudoscalar condensates are O($\epsilon$) and O($\epsilon^{3/2}$) respectively as for the Shei bound state.
ii) $\gamma \gg 1$, i.e., $\pi$ and $\sigma$ masses are comparable and of O($m$), cf. the 
exact relationship \cite{C22}
\begin{equation}
\gamma=\frac{1}{\sqrt{\eta-1}}\arctan \frac{1}{\sqrt{\eta-1}},\quad \eta=\frac{4m^2}{m_{\pi}^2}.
\label{p70}
\end{equation} 

In case i) we can neglect $\gamma$ in the scalar effective coupling multiplying $(\bar{\psi}\psi)^2$ since
it is of higher order than O($\epsilon^2$). The same diagrams contribute to the pseudoscalar 
terms as before, except that the pion proapagator becomes massive everywhere,
$\square \to \square +m_{\pi}^2$. The cancellation in the $\pi \pi$ scattering amplitude 
illustrated in Fig.~9 is upset by the bare mass term, as expected from chiral perturbation theory and low
energy theorems. However, since the correction gets an additional factor of $\gamma$, these
terms can again be ignored to LO.
As a result, the non-local effective action (\ref{p53}) of the massless NJL$_2$ model now gets 
replaced by 
\begin{eqnarray}
{\cal L}_{\rm eff} &=& \bar{\psi}({\rm i}\partial \!\!\!/-m)\psi+ \frac{\pi}{2N}(\bar{\psi}\psi)^2
\label{p71} \\
&+& \frac{2\pi m^2}{N}\bar{\psi}{\rm i}\gamma_5 \psi \frac{1}{\square + m_{\pi}^2 - \frac{4\pi m}{N}\bar{\psi}\psi}\bar{\psi}{\rm i}\gamma_5\psi,
\nonumber
\end{eqnarray}
with $m_{\pi}^2$ defined in Eq.~(\ref{p69}).
This can again be converted into a local effective action by introducing an elementary pseudoscalar
field $\Pi$. The only modification as compared to Eq.~(\ref{p54}) is the appearance of a mass term for $\Pi$,
\begin{eqnarray}
{\cal L}_{\rm eff}^{\rm loc} &=& \bar{\psi}({\rm i}\partial \!\!\!/-m)\psi+ \frac{\pi}{2N}(\bar{\psi}\psi)^2+ 
\frac{1}{2} \partial_{\mu}\Pi \partial^{\mu} \Pi
\label{p72} \\
& + & \sqrt{\frac{4\pi}{N}}m \bar{\psi}{\rm i}\gamma_5 \psi \Pi
+ \frac{2\pi m}{N}\bar{\psi}\psi \Pi^2- \frac{1}{2}m_{\pi}^2 \Pi^2.
\nonumber
\end{eqnarray}
This opens up the possibility to extend Shei's bound state to finite bare fermion masses. Of course one would then
have to verify that our assumptions concerning the order of $\epsilon$ of the valence condensates are fulfilled. 

In case ii) we go into a regime where both $\pi$ and $\sigma$ propagators may be treated as point-like. The scalar and pseudoscalar effective 
couplings become equal (to LO),
\begin{equation}
g_{\rm eff}^2 = G_{\rm eff}^2 = \frac{\pi}{N\gamma}, \quad \gamma \gg 1.
\label{p73}
\end{equation}
All higher order corrections are suppressed by inverse powers of $\gamma$, and the LO effective
Lagrangian assumes the same structure as the original Lagrangian (apart from the mass term and the replacement of $g^2$
by the effective coupling),
\begin{equation}
{\cal L}_{\rm eff} = \bar{\psi}\left({\rm i}\partial \!\!\!/ -m \right)\psi + \frac{\pi}{2N\gamma}\left[ (\bar{\psi}\psi)^2
+(\bar{\psi}{\rm i}\gamma_5\psi)^2\right].
\label{p74}
\end{equation}
In this regime, the LO effective Lagrangian of the GN$_2$ model  reduces to
\begin{equation}
{\cal L}_{\rm eff} = \bar{\psi}\left({\rm i}\partial \!\!\!/ -m \right)\psi + \frac{\pi}{2N\gamma}(\bar{\psi}\psi)^2.
\label{p75}
\end{equation}
These last two equations may be considered as the extreme heavy-fermion (or non-relativistic) limit of the original field theories.

Finally, we comment on some early work which is relevant in this context. In the 70's, there was some debate about 
the use of ``classical calculations" for fermions in field theory. These were mean field
calculations without Dirac sea, solving non-linear Dirac-HF equations \cite{C9,C10} exactly as in our ``no-sea"
effective theory. What was done at that time simply amounted to use the original Lagrangian with
a mass term added and replacing the bare coupling constant by an effective one. Shei \cite{C6} noted
that this type of calculation gave reasonable results for the GN$_2$ model, but was completely off in the case of the NJL$_2$ model.
We can now understand the reasons behind this observation, using the framework of effective field theory. In the GN$_2$ model,
the lowest order effective Lagrangian happens to agree with the ad-hoc prescription used in Refs.~\cite{C9,C10}. 
This explains why it gave useful results, at least for small filling fraction.
Incidentally, another example of a ``no-sea" calculation for this model can be found in Ref.~\cite{C7} where the effective
coupling constant $g_{\rm eff}^2=\pi/N$ was determined ``phenomenologically", see Eq.~(4.17) of that paper.
We have now derived this number microscopically and are able to improve the calculation in a systematic fashion.
In the NJL$_2$ model, the naive recipe for writing down the effective action breaks down
due to the pion pole. One cannot ignore
the singular momentum dependence of the effective pseudoscalar coupling ($\sim 1/k^2$) which gives rise to long range non-localities.
This is the reason for the observed discrepancy between short range ``no-sea" and long range semi-classical mean fields.
It is amusing that in the heavy-fermion limit, we do recover an effective action which agrees with the simple recipe
of adding a mass term and introducing an effective coupling, see Eq.~(\ref{p74}).
Hence the technically rather nice classical calculations done in the 70's for the chiral limit of the NJL$_2$ model (where they had to fail,
as we now understand) can be ``recycled" to solve the NJL$_2$ model in the heavy-fermion limit $\gamma \gg 1$.
In this regime, the approximations can be justified by effective field theory.
In the representation of the $\gamma$-matrices (\ref{p1a}), the bound state
found by Lee et al. \cite{C10} has the spinor wave function (filling fraction $\nu$, energy $E_0$) 
\begin{eqnarray}
\psi_0 &=&   \sqrt{\frac{(m-E_0)}{2G}}\frac{1}{\cosh^2 \xi+ \alpha^2 \sinh^2 \xi}
\nonumber \\
& \times & \left( \begin{array}{c}
\alpha \sinh \xi + \cosh \xi \\ \alpha \sinh \xi - \cosh \xi \end{array} \right),
\nonumber \\
G & = & \frac{\pi \nu}{2\gamma}, \quad \alpha = \tan \frac{G}{2}, 
\nonumber \\
\quad E_0 & = & m\cos G, \quad \xi = mx \sin G.
\label{p76}
\end{eqnarray}
We have checked that it satisfies the (classical) non-linear Dirac equation obtained from the effective Lagrangian (\ref{p74})
exactly. The baryon mass is given by 
\begin{equation}
M_B= nm \frac{\sin G}{G} = nm \left(1- \frac{\pi^2 \nu^2}{24 \gamma^2} + ... \right)
\label{p77}
\end{equation}
Due to the truncation of the effective action, only the two listed terms of the series expansion can be trusted.
At the same time the induced fermion density gets suppressed by a factor of $1/\gamma$ as compared to the
valence density.
The corresponding result for the GN$_2$ model with effective action (\ref{p75}) can be read off Eq.~(\ref{p39})
(for $\gamma \gg 1$) and agrees with (\ref{p77}) to this order in $\nu$. As a matter of fact, the spinors $\psi_0$
also become equal in the limit $\gamma \to \infty$, as can be seen by expanding in powers of $\nu$. The reason
can be traced back to the fact that the ratio of pseudoscalar to scalar condensates decreases like $1/\gamma$.
Thus we learn that in the heavy-fermion limit, the difference between discrete and continuous chiral symmetry
of the four-fermion interaction becomes less and less important, and 
baryons of the GN$_2$ and NJL$_2$ model approach each other.
As far as the NJL$_2$ model is concerned, there is so far no independent information available about this regime
these findings could be compared to.
\section{Summary and conclusions}
Using exactly solvable four-fermion models as theoretical laboratory, we have demonstrated that ``integrating 
out the Dirac sea" is indeed a viable concept. Tractable effective Lagrangians could be derived, valid for 
multi-fermion bound states with weakly occupied valence level or large bare fermion masses. Whereas the
original semi-classical calculations of bound states in GN$_2$ and NJL$_2$ models required sophisticated
and highly specialized techniques to deal with the Dirac sea, the effective theories amount to solving the classical 
Euler-Lagrange equation, here a non-linear Dirac equation. One might suspect that the difficulties of the full
non-perturbative calculations come back in the derivation of the effective Lagrangian, but this is not quite so.
Although this derivation has non-perturbative features in the form of unavoidable resummations, the actual 
calculation is based on standard one-loop Feynman diagrams and thus perturbative. This suggests
that similar techniques can also be applied to more realistic, higher dimensional theories.

In the GN$_2$ case, we were able to derive the NNLO approximation in some small expansion parameter. 
Since the exact DHN baryons and their generalization to finite bare fermion mass are known, we could show that the 
resulting effective theory works quantitatively. Thus baryon binding energies can be predicted ``classically"
in the full range of filling fraction and bare mass at the 1$\%$ level, quite unexpectedly for us. 
As far as the formalism is concerned, the main lesson
we learned is that the bare parameters of the original theory ($m_0,g^2$) disappear owing to resummations. It would not 
make sense to truncate the procedure at any fixed number of loops. All questions of renormalization and UV
divergences are then properly dealt with and do not show up any more in the effective theory, which is of
course finite.

In the NJL$_2$ case, when applying the same strategy we ran into a new difficulty. The pseudoscalar effective
coupling develops a singularity at $k^2=0$ due to the massless pion pole. Although one can again
identify the LO of a systematic expansion in a small parameter, this necessarily causes a long range non-locality
of the effective Lagrangian. The purely fermionic, non-local effective theory can be cast into a more convenient 
``almost local" form by introducing an additional elementary pion field. The resulting theory contains elementary fermions
and pions but
is fully consistent, judging from the derivation and the comparison with Shei's bound states. The role of the massless pion explains
the conspicuous difference between short range mean fields in the GN$_2$ model and long range mean fields
in the NJL$_2$ model, which had caused headaches in early, more naive attempts to treat fermions classically.
Moreover, we could show how the important phenomenon of ``induced fermion number" can be recovered 
in a ``no-sea" effective theory.
 
Finally, notice that no attempt was made to reduce the Dirac equation to a non-relativistic Schr\"odinger equation.
Although this would be possible, we feel that it would only make our formulae more messy. Our main focus
here was not on relativistic kinematics, but on the dynamical effects of the Dirac sea.

\end{document}